\documentclass[]{WileyMSP-template}
\usepackage{amsmath,amssymb,amsfonts,amsthm,epsfig}
\usepackage[applemac]{inputenc}
\usepackage[all]{xy}

\def \ep{\varepsilon}
\def \bbC{\mathbb{C}}
\def \bbR{\mathbb{R}}

\def \calR{\mathcal{R}}

\def \calM{\mathcal{M}}
\def \bq{\begin{equation}}
\def \eq{\end{equation}}
\def \ba{\begin{array}}
\def \ea{\end{array}}

\DeclareMathOperator{\Tr}{Tr}

\begin{document}

\pagestyle{fancy}
\rhead{\includegraphics[width=2.5cm]{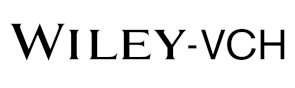}}

\title{Characterizing the topological properties of one-dimensional non-hermitian systems without the Berry-Zak phase}
\maketitle

\author{D. Felbacq}
\author{E. Rousseau}

\dedication{}
\begin{affiliations}

Prof. D. Felbacq\\
L2C, CNRS, Univ Montpellier, 34095 Montpellier, France \\
Email Address: didier.felbacq@umontpellier.fr

Dr. E. Rousseau\\
L2C, CNRS, Univ Montpellier, 34095 Montpellier, France\\
\end{affiliations}


\begin{abstract}
A new method is proposed to predict the topological properties of one-dimensional periodic structures in wave physics, including quantum mechanics. From Bloch waves, a unique complex valued function is constructed, exhibiting poles and zeros. The sequence of poles and zeros of this function is a topological invariant that can be linked to the Berry-Zak phase. Since the characterization of the topological properties is done in the complex plane, it can easily be extended to the case of non-hermitian systems. The sequence of poles and zeros allows to predict topological phase transitions.
\end{abstract}

\section{Introduction}
A considerable amount of work has been devoted to the study of the topological properties of photonic structures \cite{topophot}. The word topological means that what is at stake are the properties of a structure that are stable under a continuous variation of the parameters defining the it. For instance, the existence of a band gap is a topological property since a not too large variation of the properties (e.g. the size of the basic cell, the values of the electromagnetic parameters) do not close the gap. In some cases, the topological properties can be characterized by an integer number, a quantity that obviously remains constant over small continuous variations \cite{colloquium}. Of course, for larger variations it may happen that, e.g., the gap closes, which can lead to a change into the integer number. This will be called a topological transition. First attempts to find topological properties in photonic structures were made by mimicking the field of topological insulators \cite{soljacic}: the time-reversal invariance was broken by the use of gyromagnetic materials controled by a magnetic field. The devices considered there are quite complicated and specific.
 A major breakthrough was made when it was realized that topological effects could be obtained in purely dielectric structures \cite{scheme}. As a matter of fact, topological effect can be obtained for very simple structures: one dimensional stratified media exhibit boundary modes that are topologically protected \cite{Chan2014}. These can be analyzed using theoretical tools that were developed a long time ago \cite{kohn} and which have been given a second look in the context of geometric phases: in \cite{zak}, it was shown that the properties exhibited by Kohn could be interpreted using the mathematical apparatus developed by \cite{simon} in view of the results obtained by Berry in \cite{berry}. Mathematically speaking, this comes under the domain of vector bundles endowed with a connection \cite{nash,nakahara}. Simon introduced in \cite{simon} the now celebrated connection 1-form $A(k)=\langle u_k,\nabla_k u_k \rangle$, where $u_k$ is a Bloch mode. This connection is often called the Berry connection, while it is in fact a specific case of the Levi-Civita connection \cite{dewitt}. Zak's article was probably the first work applying the concept of geometric phase to Bloch waves. Recently, there was an interest in the possibility of extending these results to non-hermitian Hamiltonians \cite{okuma, lieu, shen, kawabata} and, in the context of photonic crystals, to media with losses \cite{mario}. All the preceding results were obtained by generalizing the Levi-Civita connection for a non-hermitian bundle. In the present work, we propose a new approach to the topological properties of 1D periodic structures. We show that the topological properties can be analyzed without reference to the Levi-Civita-Simon-Berry connection. We introduce a function of the wavenumber that presents poles and zeros. The arrangement of the poles and zeros characterizes the topological properties of the medium. It turns out that this pole-zero structure extends naturally to the situation when losses are present in the materials out of which the structure is made, i.e. to the non-hermitian situation.
In the first section, we recall the elements of the theory of wave propagation in 1D structures, comprising Bloch waves. In the second section, we develop our approach and introduce the function that will prove to be a clue to the understanding of the topological properties. In the third section, we make the link with the usual approach using the geometrical phase when the medium under consideration are lossless. Finally, we extend the approach to the situation where loss is added. Throughout the sections, numerical illustrations are provided in order to clarify the somewhat abstract statements.
\section{Wave propagation in a one dimensional structure}
\begin{figure}
	\begin{center}
		\includegraphics[width=8cm]{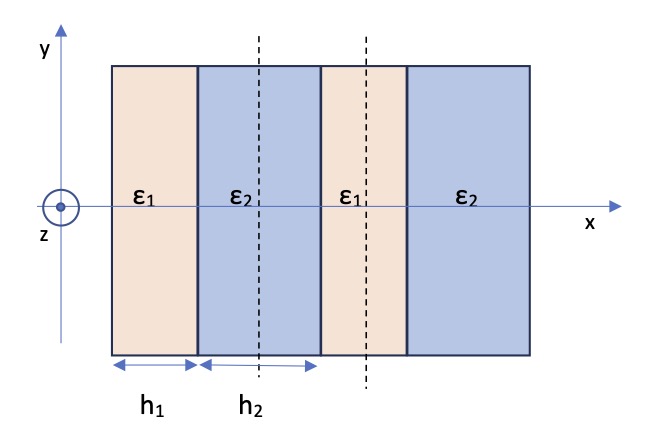}
		\caption{An example of 1D structure. It is a stratified photonic crystal with two homogeneous slabs in the basic cell. The vertical dashed lines correspond to the two origins for which the potential is symmetric.}
		\label{paramphc}
	\end{center}
\end{figure}

An example of a 1D medium is depicted in fig. (\ref{paramphc}). We have chosen to represent a 1D stratified photonic crystal, but the results that we obtain apply to continously varying structure as well as to acoustic structure and to quantum waves in a 1D potential. 
For definiteness, we proceed by using the vocabulary of electromagnetism in the following. However, we will use occasionally the word "potential" generically to designate the permittivity (permeability) or the quantum confining potential. With an abuse of notation, we will talk of the "potential $V(x)$" to denote generically these quantities. We will further make the hypothesis that there is an inversion symmetry in the medium, that is, an origin can be chosen in such a way that $V(x)=V(-x)$.

Let us briefly recall the theory of 1D media \cite{felbacq1998}. We consider time-harmonic fields (time-dependence of $e^{-i\omega t}$) that are invariant along the axis $y$ and $z$ and depend only on the variable $x$ (see fig. (\ref{paramphc}) ).
Since the medium under consideration is invariant along two directions of space, the electromagnetic field can be decomposed as a sum of two linearly polarized fields. Either the electric field is linearly polarized along $z$ ($E_{||}$ case) or the magnetic field is linearly polarized along $z$ ($H_{||}$ case). In both cases, we denote by $u(x)$ the function representing the field, i.e. $E_z(x)=u(x) e_z$  ($E_{||}$ case) or $H_z(x)=u(x) e_z$ ($H_{||}$ case).
The medium is described by a periodic relative permittivity $\ep(x)$ and a periodic relative permeability $\mu(x)$. The system of units is chosen in such a way that $\ep(x+1)=\ep(x),\, \mu(x+1)=\mu(x)$.
From the Maxwell system, the following equation is obtained, valid in the Schwartz distributions meaning:
\bq
Hu=k_0^2u,
\eq
we denote $k_0=\omega/c$ the wavenumber, and
\bq
H=-p^{-1}(x) \frac{d}{dx}\left( q^{-1}(x)\frac{d .}{dx} \right),
\eq
where, according to the polarization:
\bq
E_{||}:q(x)=\mu(x),\, p(x)=\ep(x) , \, H_{||}: q(x)=\ep(x),\, p(x)=\mu(x).
\eq

This equation is more conveniently rewritten as an order one differential system:
\bq
\frac{d}{dx} U=
\left( \ba {cc}0 & 1\\ -k^2_0 p(x)  & 0 \ea\right) U,\eq
where $$U=\left(\ba{r}u \\ \frac{1}{q(x)}\frac{du}{dx} \ea\right).$$ 
In the following, we denote \bq u'=\frac{1}{q(x)}\frac{du}{dx}.\eq

From the general theory of ordinary differential equation, there exists a so-called resolvent matrix $\mathcal{R}(x,y)$ such that $U(x)=\mathcal{R}(x,y) U(y)$.

Over one period, the values of $U(1)$ and $U(0)$ are related by the so-called monodromy matrix: \bq\calM(k_0)=\calR(1,0).\eq
This matrix is unimodular (i.e. $\det \calM=1$) and it characterizes the band structure in the hermitian case. The monodromy matrix depends on the norm of the wavevector in vacuum (or on the energy of the system in case of quantum physics). The characteristic polynomial of $\calM$ reads as: $X^2-tr(\calM)X+1$, therefore three sets can be defined according to the nature of the eigenvalues of $\calM$ \cite{felbacq2003,dofresnel}:
\begin{itemize}
	\item $G=\left\{k_0 \in \bbR, \, |\Tr(\calM(k_0))|> 2\right\}$, for which the eigenvalues are real and inverse one of the other. This corresponds to non progagative modes, i.e. band gaps.
	\item $B=\left\{k_0 \in \bbR, \, |\Tr(\calM(k_0))|< 2\right\}$, for which the eigenvalues are complex of modulus one and conjugated. This corresponds to progagative modes, i.e. conduction bands. The eigenvalues can be written $e^{\pm i \theta}$, with $\theta \in [-\pi,+\pi]$ the Bloch number and the interval $[-\pi,+\pi]$ is the so-called Brillouin zone.
	\item $\Delta=\left\{k_0 \in \bbR, \, |\Tr(\calM(k_0))|= 2\right\}$, for which the eigenvalues are $\pm 1$. The eigenvalues are of multiplicity 2. We denote $\Delta_0$ the subset of $\Delta$ for which $\calM(k_0)=\pm I_d$.
\end{itemize}
Conventionaly, Bloch waves are associated to eigenvalues of modulus one and therefore to propagative modes in the structure. This corresponds to the set $B$. Looking at the various sets, we see that the distinction between the various domain is purely qualitative: eigenmodes always exist in the system but they are unbounded for $k_0 \in G$. That is why we call "generalized Bloch modes" the modes corresponding to the sets $G$ and $\Delta$. In the band gaps, the solutions to the wave equation are not bounded over the infinite structure, but they play a crucial role in the case of a finite or semi-infinite medium. We therefore take as parameter the wavenumber $k_0$ that will vary in $\bbR^+$ and study the evolution of eigenvalues and eigenvectors with respect to $k_0$.

For $k_0 \in B$, the fields that can exist in the structure are superposition of Bloch waves $\psi(x;\theta,k_0)$. The Bloch wave are quasi-periodic in the $x$ variable, that is, there are of the form: ${\psi(x;\theta,k_0)=e^{i\theta x}u(x;\theta,k_0)}$, where $u(x;\theta,k_0)$ is 1-periodic in the $x$ variable: ${u(x+1;\theta,k_0)=u(x;\theta,k_0)}$.


It is convenient to transform the Brillouin zone into the circle $S^1=\{z \in \bbC,\, |z|=1\}$, that is, we associate to $\theta \in [-\pi,\pi]$ the complex number $z=e^{i\theta} \in S^1$. Any function defined on the interval $[-\pi,\pi]$  can be considered as a function on $S^1$. From now on the Bloch modes are thus denoted $u(x;z,k_0)$. For each value of $z$, there corresponds a set of wavenumbers: \bq \label{branch} k_{0,j}(z), \, j=1\hdots, \eq where $k_{0,j}^2$ are the eigenvalues of $H$ and for each of these wavenumbers a complex vector space of dimension 1. The collection of these vector spaces constitutes the Bloch bundle associated with the branch $k_{0,j}(z)$. 
As a function of $\theta$, it is not obvious that $u(x;\theta,k_0)$ satisfies the condition that $u(x;\pi,k_0)=u(x;-\pi,k_0)$. However, it is a general result that for any one dimensional complex bundle over $S^1$, there exists a \textit {continuous} function $z \in S^1 \to u(x;z,k_0)$ \cite{nash} (such a function is a called a "section" of the bundle). This does not mean that no topological effect is to be expected as we will see below, because we have the additional hypothesis that the potential is symmetric.

Due to the fact that the Bloch waves depend on both the direct and indirect variables $x$ and $\theta$ this representation of the Bloch bundle is not easy to handle . We present another, simpler, representation of the bundle.

In order to make the discussion less arid, we shall give a numerical illustration of the concept that we deal with (the code written in Matlab is available in the supplementary material). We consider the case of a binary medium, where the period is made of two homogeneous layers of relative permittivity $\ep_1$ and $\ep_2$ and width $d_1$ and $d_2$ (see fig. (\ref{paramphc})). Let us denote $\nu_j=\sqrt{\ep_j},\, j=1,2$.

\begin{figure}
	\begin{center}
		\includegraphics[width=8cm]{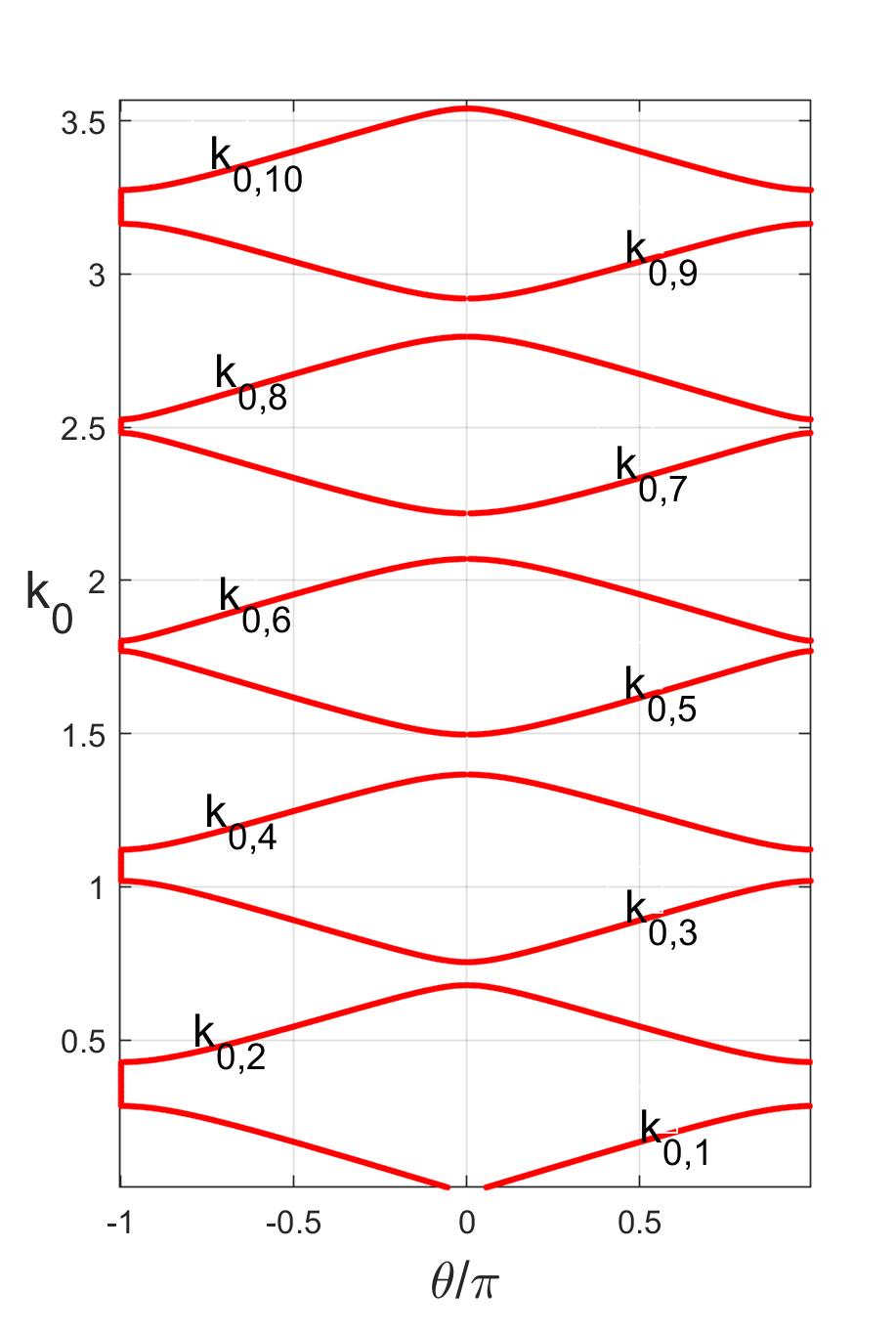}
		\caption{band structure for a photonic crystal with parameters $\ep_1=3.8,\,\ep_2=1$, $h_1=0.42,\,h_2=0.58$ and the permeabilities are equal to $1$. These are the values used in \cite{Chan2014}. The labeling of the branches is as in (\ref{branch}).}
		\label{band1d}
	\end{center}
\end{figure}

For each layer, the monodromy matrix has the form
\bq
\mathcal{M}_j=
\left(
\ba {cc}
\cos(k_0 \nu_j d_j) & \frac{1}{k_0 \nu_j} \sin(k_0 \nu_j d_j) \\
-k_0 \nu_j\sin(k_0 \nu_j d_j) & \cos(k_0 \nu_j d_j)
\ea
\right),
\eq
and the complete monodromy matrix is simply $\mathcal{M}=\mathcal{M}_2\mathcal{M}_1$.
The dispersion relation is obtained by computing the trace of $\mathcal{M}$, which leads to the equation
\bq
\cos(\theta)= \cos(k_0 \nu_1 d_1)\cos(k_0 \nu_2 d_2)-\frac{1}{2}\left( \frac{\nu_2}{\nu_1}+\frac{\nu_1}{\nu_2}\right)\sin(k_0 \nu_1 d_1)\sin(k_0 \nu_2 d_2)
\eq
and the eigenvectors $U^{\pm}$ are obtained by diagonalizing the monodromy matrix $\calM$.
An example of a band structure is given in fig. (\ref{band1d}) where the conduction bands are labeled as in (\ref{branch}).

Let us now extend this setting to complex values of the energies. Consider the Bloch variety:
$$
F=\left\{(z,k_0) \in \bbC^2,\, \exists \, U\neq 0: \calM(k_0) U=z U \right \}
$$
The Bloch variety is obtained explicitly by the solving the characteristic polynomial of $\calM$:
$$
\det(\calM(k_0)-z I_2)=0.
$$
This allows to consider the situation when the potential is not necessarily real (i.e. the case of a non-hermitian quantum system or of media with losses). The Bloch variety defines the zeros set of an analytic function of the two variable $(k_0,z)$. In the community of integrable systems, it is called the spectral curve of the system \cite{babelon}. The vector bundle envisioned previously extends to a vector bundle over this curve. 
The situation of a real potential and real energies corresponds to the set $F_0 \subset F$ defined by
$$
F_0=\left\{(z,k_0) \in \bbC \times \bbR,\, \exists \, U \neq 0: \calM(k_0) U=z U \right \}.
$$


Generically, the monodromy matrix can be put in diagonal form, however for $(k_0,z) \in \Delta \setminus \Delta_0$, a particular situation happens: the eigenvalue $\pm 1$ has a multiplicity of $2$ but the eigenspace is of dimension $1$. The points where this happens are called ramification points.
The set $\Delta_0$ where $\calM=\pm I_2$ corresponds to a non-generic situation that is a topological phase transition, as will be shown in the following. Indeed, this corresponds to two bands touching at one point $z=\pm 1$. This singularity can be removed by an infinitesimal variation of the parameters. It separates two topological phases stable under small variations of the parameters far enough from this singularity.

\section{Topological characterization using poles and zeros}
Let us consider for now the usual case of a real potential or lossless media. The eigenvectors $U^{\pm}(x_0)$ of $\calM$ are the boundary values of the generalized Bloch waves at an origin $x_0$ that can be chosen at will. Changing the value of $x_0$ amounts to changing the basic cell to $[x_0,x_0+1]$. It is equivalent to a change of gauge in real space whenever an infinite medium is considered. When finite structures are considered different choices of the origin will correspond to different physical properties as shall be seen later on.

The eigenvectors  $U^{\pm}$ are complex conjugate when the potential is real and for $k_0 \in B$. Using the resolvent matrix, the value of a Bloch mode at any point $x$ in the period is obtained by using the relation: $U^{\pm}(x)=\calR(x,x_0)U^{\pm}(x_0)$. 
From these considerations, we conclude that the Bloch eigenspace at a point $(k_0,z)$ is entirely determined by the eigenvectors $U^{\pm}(x_0)$. Therefore the Bloch bundle, i.e. the collection of all the eigenspaces as $z$ describes $S^1$ is isomorphic to the vector bundle of eigenvectors of the monodromy matrix $\calM$. As we have already said, it is a complex vector bundle over $S^1$, and therefore it is trivial, which means that there exists a non-vanishing section, that is, a continuous parametrization of a Bloch mode all around the Brillouin zone. We recover here one of the conclusions of \cite{zak}. Still, there can be topological properties provided that we take into account the hypothesis that the potential has an inversion symmetry $\sigma$: $V(x)=V(-x)$

Two different origins $x_0$ and $x_1$ can be chosen such that $V(x)=V(-x)$. These two points are such that $x_1-x_0=1/2 \mod(1)$ (see fig. (\ref{paramphc}) where the two origins are indicated as dashed lines). Let us assume that the boundary values are chosen at one of the two points such that $V(x)=V(-x)$, that is, we fix the gauge in real space. This point is now the new origin $x=0$.

For the eigenvectors $U^{\pm}$ of the monodromy matrix $\calM$, the inversion symmetry $\sigma$ acts as $\sigma(U^{\pm})=\sigma_z U^{\pm}$, where $\sigma_z$ is the Pauli matrix $\begin{pmatrix}1 & 0\\ 0 & -1 \end{pmatrix}$. This is so because under the change $x \rightarrow -x$, the derivative changes sign and the wave propagates backwards: the corresponding monodromy matrix is $\calM^{-1}$. The inversion symmetry implies that if U is an eigenvector of the monodromy matrix with the eigenvalue $z$ then  $\sigma_z U$ is an eigenvector of the monodromy matrix but with the eigenvalue $1/z$. As a consequence, we have the following result:

\textit{
When $V(x)=V(-x)$, it holds:
\bq \label{sigmabasis}
\sigma_z \calM \sigma_z=\calM^{-1}, \hbox{ and the basis of eigenvectors is of the form } (U,\sigma_z U). 
\eq
}

For $k_0 \in B$, it holds, projectively, $U^*=\sigma_z U$ ($*$ denotes complex conjugaison). When we say that two vectors are equal ``projectively'', we mean that they are colinear. Given a vector $U=(u(0;k_0,z),u'(0;k_0,z))^t$, the vector space generated by $U$ is denoted $\tilde{\chi}=[u(0;k_0,z):u'(0;k_0,z)]$. This is the standard notation for an element of the projective space $\bbC P_1$ which is the set of all complex lines going through the origin in the complex plane $\bbC^2$. This space is equivalent to the Riemann sphere, that is $\bbC$ together with a point $\infty$ at infinity.

This suggests to find a function that characterizes an eigenspace. Given an eigenvector $U=(u(0;k_0,z),u'(0;k_0,z))^t$, we define \bq \chi(k_0,z)=u(0;k_0,z)/u'(0;k_0,z). \eq This quantity does not depend upon the specific eigenvector that is chosen to represent the eigenspace, i.e. $U$ and $\gamma U,\, \gamma \in \bbC$ define the same function $\chi$. 

For $k_0 \in B$, the ratio $\chi(k_0,z)=u(0;k_0,z)/u'(0;k_0,z)$ satisfies the relation $\chi^*(k_0,z)=-\chi(k_0,z)$ and therefore, thanks to (\ref{sigmabasis}), it is purely imaginary in the conduction bands for a real potential. 

For a given wavenumber $k_0$, there are two eigenvectors $(U^+=U, U^-=\sigma_z U)$ with eigenvalues $z$ and $1/z$ respectively. Therefore two functions $\chi^{\pm}$ obtained from the components of the eigenvector $U^+$ or $U^-$, respectively. These functions satisfy therefore the relation:
\bq
\chi^+(k_0,z)=-\chi^-(k_0,1/z),
\eq
%
\begin{figure}
	\begin{center}
		\includegraphics[width=12cm]{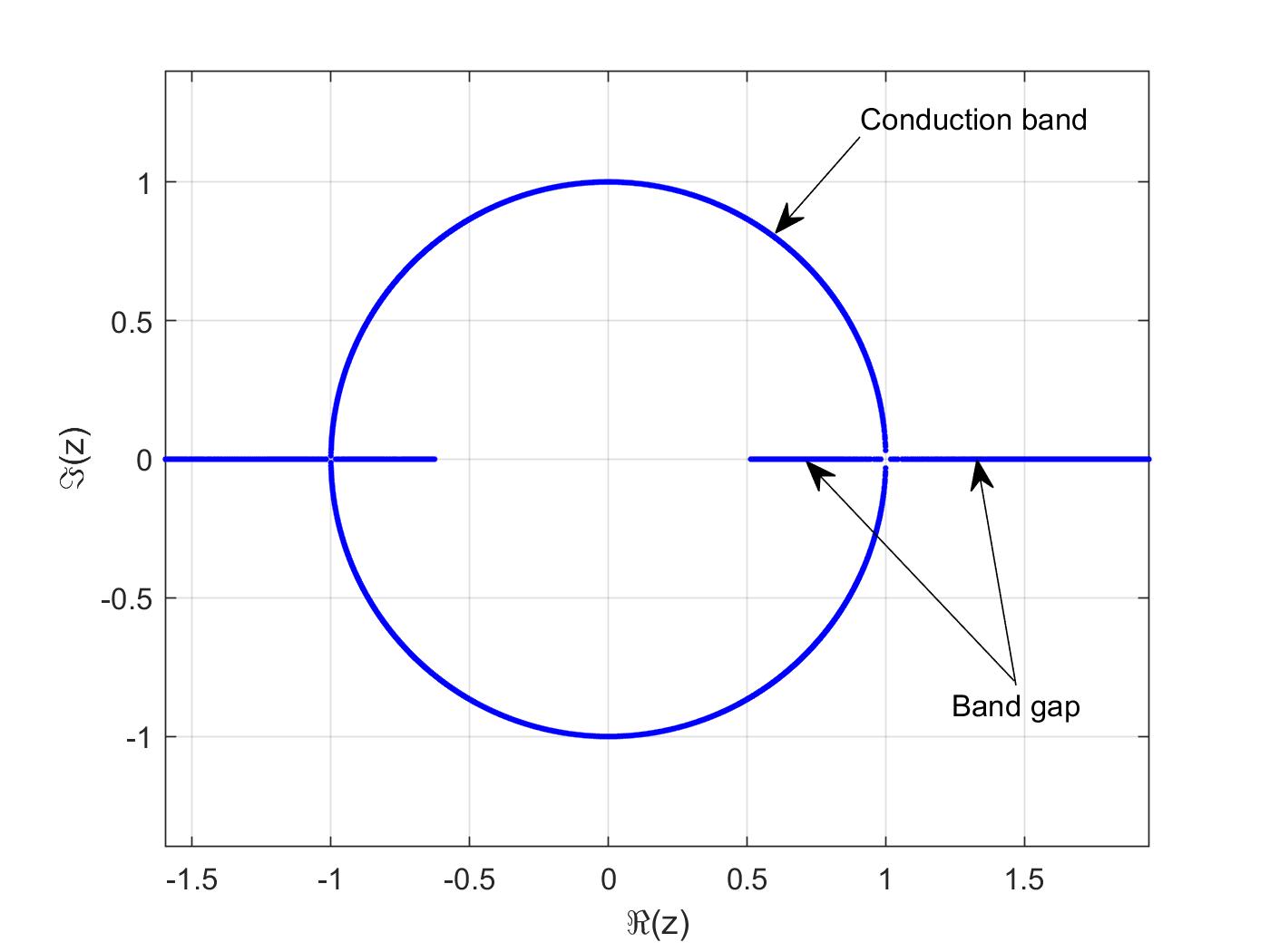}
		\caption{Eigenvalues of the monodromy matrix in the complex plane. The circle corresponds to $B$ and the real intervals to $G$. The transitions between $G$ and $B$, i.e. the set $\Delta$, correspond to $z=\pm 1$.\label{eigvalueplane}}
	\end{center}
\end{figure}
 Let us show that these can be combined to provide a single function.

\begin{figure}
	\begin{center}
		\includegraphics[width=8cm]{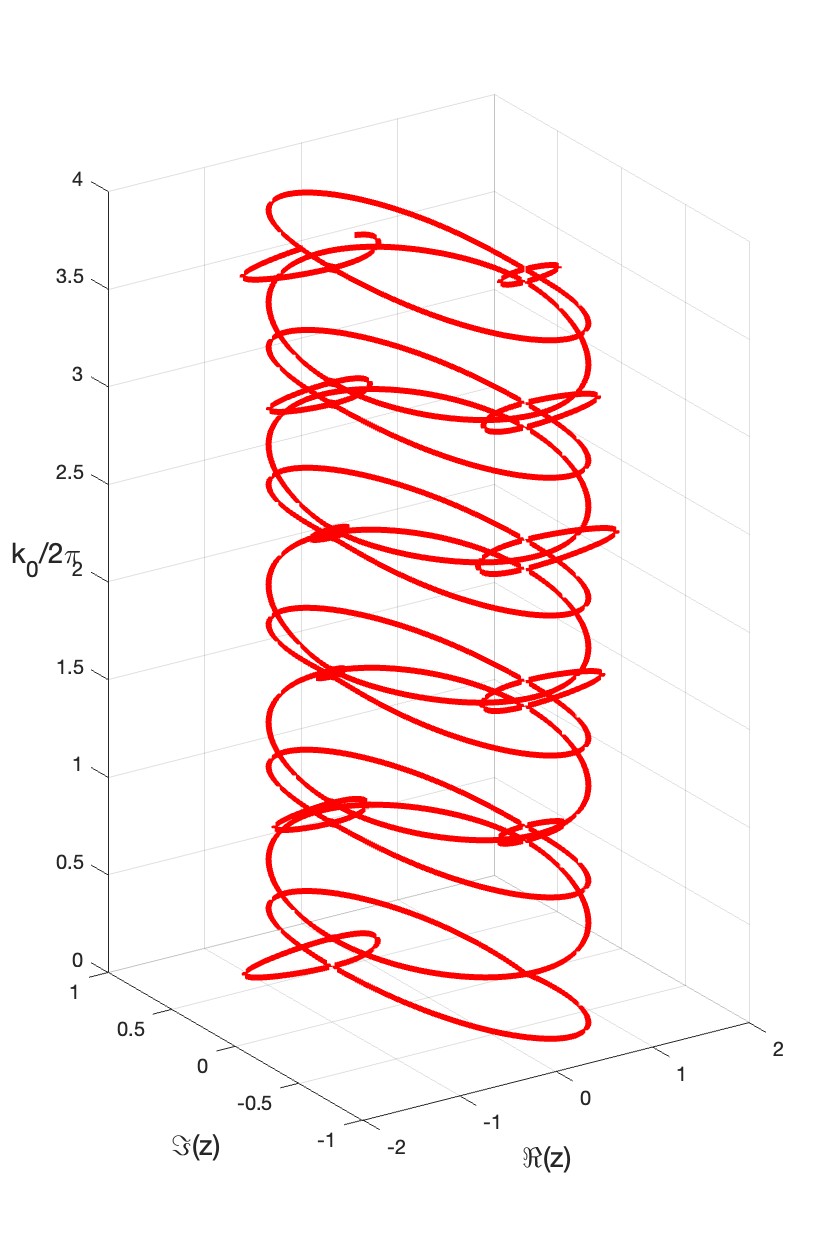}
		\caption{Evolution of the eigenvalues $z$ as functions of $k_0$. The band gaps correspond to the (real) ovoid regions and the conduction bands to the helical parts of the curve. \label{eigvalblowup}}
	\end{center}
\end{figure}

For $k_0 \in  G$, since one of the numbers $(|z|,1/|z|)$ is lower than $1$, we can define a single function $\chi$ such that:
\bq
\chi(k_0)=\left\{ 
\ba{lr}
\chi^+(k_0,z), & |z|<1 \\ -\chi^+(k_0,z), & |z|>1
\ea
\right . .
\eq
In order to extend this definition to the conduction band, a criterium is needed to distinguish between $z$ and $1/z$. 
Let us consider the curves: $k_0 \to (z(k_0),1/z(k_0))$, which represent the evolution of the eigenvalues as functions of $k_0$. These are curves in the complex planes. This is represented in fig. (\ref{eigvalueplane}) and (\ref{eigvalblowup}). The curve in fig. (\ref{eigvalueplane}) represents the eigenvalues of the monodromy matrix in the complex plane when varying the wavenumber $k_0/2\pi$ between $0$ and $4$. This curve shows all the possible eigenvalues for a given geometry. The parameters used for the calculations are indicated in the caption of fig. (\ref{band1d}). The eigenvalues corresponding to propagating waves, i.e. corresponding to the set $B$, lie in the unit circle. The eigenvalues corresponding to  the band gap $G$ have a null imaginary part. They give rise to the line segments around $z=\pm 1$. When the potential is real, the ramification points are $1$ and $-1$. The curve in fig. (\ref{eigvalblowup}) is the "blow-up" of the preceding curve by plotting directly the skew curve $k_0 \to (z(k_0),1/z(k_0))$. 

\begin{figure}
	\begin{center}
		\includegraphics[width=8cm]{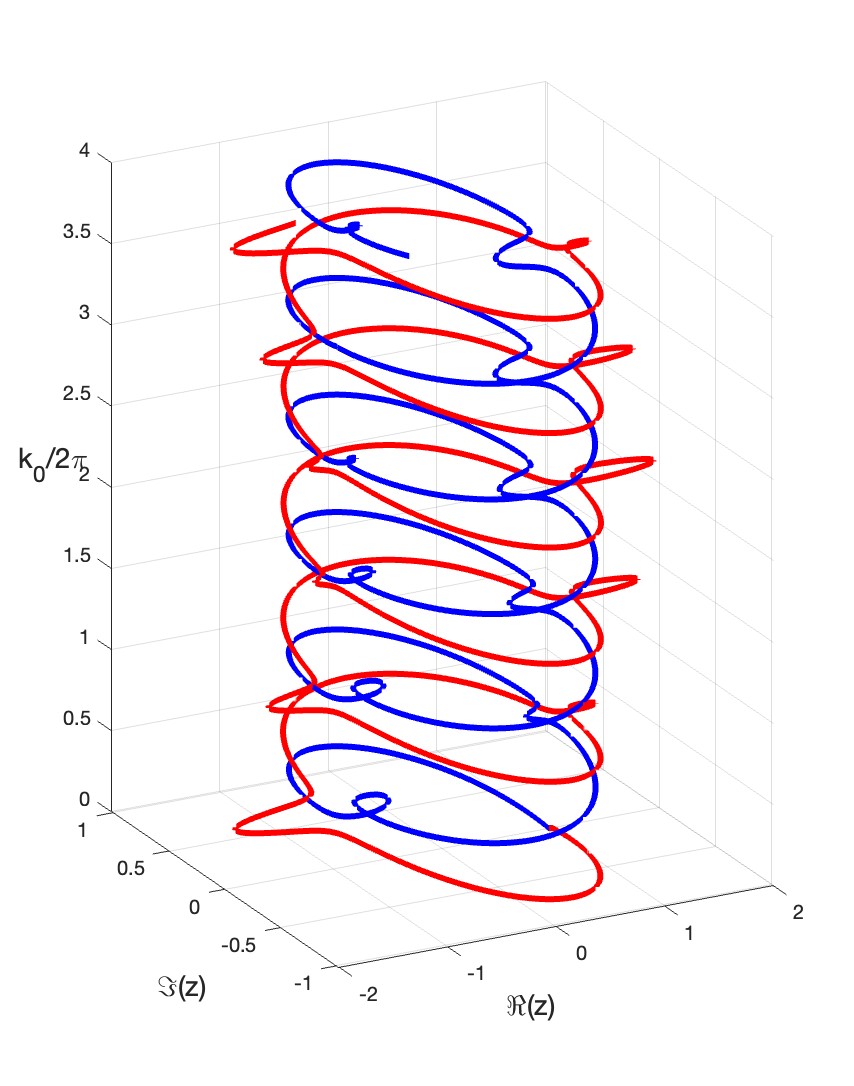}
		\caption{Evolution of the eigenvalues as functions of $k_0$ when small losses are added. The curves are now can now be distinguished (except at the bottom when $k_0 \to 0$). \label{eigvalueblowup}}
	\end{center}
\end{figure}

\medskip
At the ramification point, the curves cross at a right angle, which makes it seemingly impossible to follow one eigenvalue by continuity from $B$ to $G$. However, this degeneracy is in fact directly linked to the stringent condition that be potential be real, since it is a consequence of the fact that the monodromy has eigenvalues $\pm 1$ at the boundaries of the conduction bands. This degeneracy can be lifted by using a limiting absorption principle, that is, by adding a small imaginary part $\delta$ either to the potential or to the frequency. 
Indeed, if we replace $k_0$ by $k_0+i \delta$, the equation $tr(\calM(k_0))=\pm 2$ is replaced by the equation $tr(\calM(k_0+i\delta))=\pm 2$. Therefore the crossing points are no longer real (generically). This is what is done in fig. (\ref{eigvalueblowup}) where a small imaginary value of $10^{-2}$ was added to $k_0$. The real part of $k_0$ is used as a blow-up parameter to plot the eigenvalues as curves in $\bbR^3$. It is seen that the curves no longer cross and therefore the eigenvalues can be followed individually.

\begin{figure}
	\begin{center}
		\includegraphics[width=12cm]{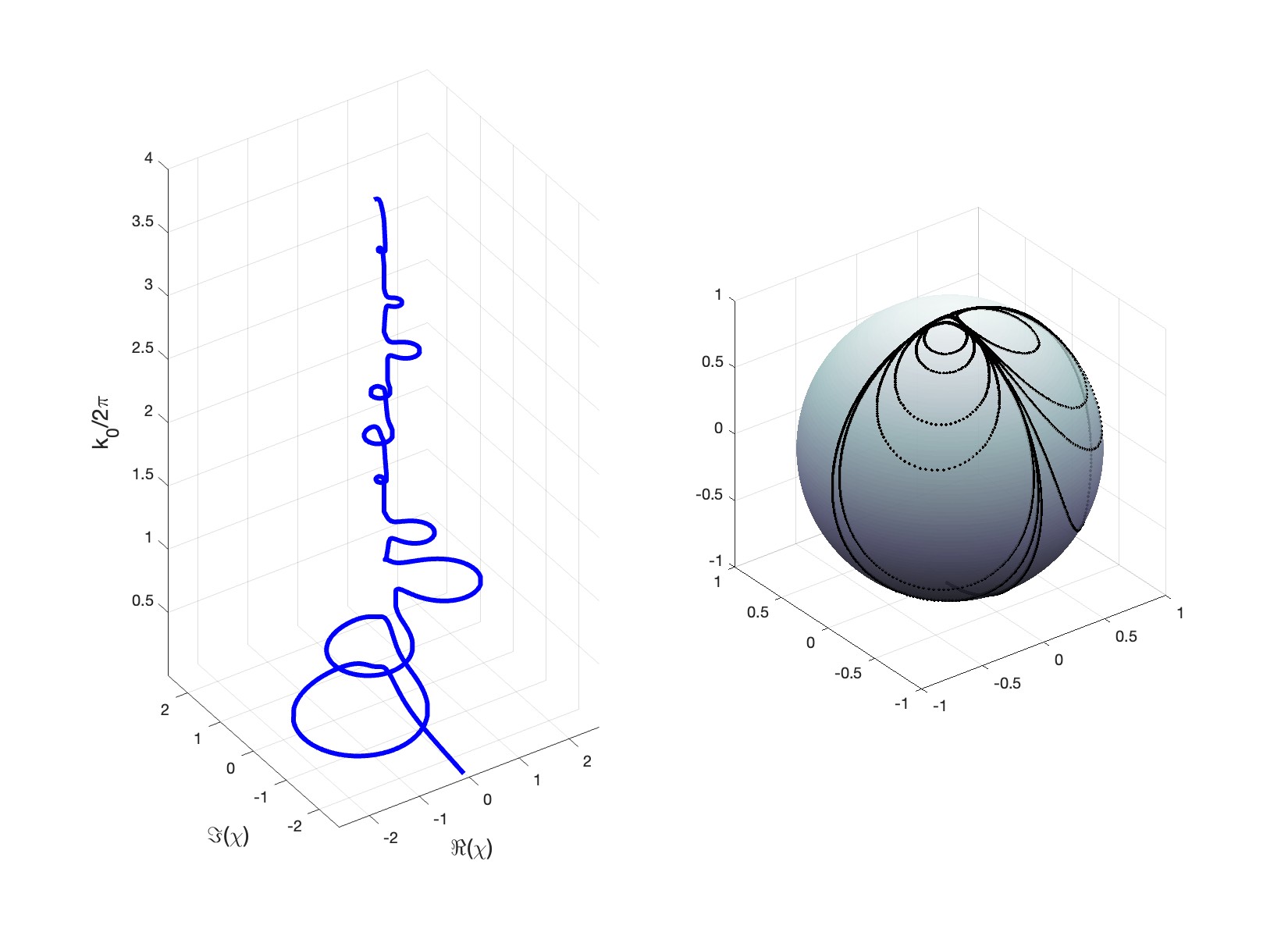}
		\caption{On the right, the graph of the function $\chi(k_0)$ on the Riemann sphere. The north pole corresponds to 0 and the south pole to infinity. On the left, the same but blowed up by using the parameter $k_0$.}
		\label{riesphere}
	\end{center}
\end{figure}

Since the eigenvalues can be distinguished, it is also possible to follow the eigenvectors by continuity, i.e. to resolve the ramification points. Therefore the same holds for the functions $\chi^{\pm}$. For these functions, the extension problem appears when there is a pole. When a small imaginary part is added to $k_0$, the poles of $\chi ^{\pm}$ gain a small imaginary part and the restriction of these functions to the real axis of $k_0$ is now continuous. It is illustrated in fig.(\ref{riesphere}) where the values of $\chi^{\pm}$ are plotted on the Riemann sphere by stereographic projection.
We conclude the following:

\textit{There is a unique function $\chi(k_0,z)$ corresponding to eigenspaces associated with eigenvalues lower than $1$ in modulus for $k_0 \in G$.}

At the heart of the topological properties captured by the function $\chi$ are the poles and zeros that it possesses. A zero is associated with the eigenvector $(0,1)^t$ and a pole to the eigenvector $(1,0)^t$.
We have the following result:

\textit{For $k_0\in B$, since $U^*(k_0,z)=U(k_0,1/z)$, we have that either $\chi(k_0,z)$ is null at $z=\pm 1$ or it has a pole.}

The values of $k_0$ for which there can be a zero or a pole are characterized as follows. Assume that there is a zero $k_Z$ in the interior of a band. Then $\chi^{\pm}$ are both null and therefore the eigenvectors of $\calM(k_Z)$ are linearly dependent and the eigenvalues $z$ and $1/z$ are equal. Consequently, $tr(\calM)=\pm 2$. If $tr(\calM)$ is not transversal there, i.e. if the derivative of $tr(\calM)$ is null, then it is a non-generic point that can be removed by an infinitesimal variation of the parameters. Similarly, if there is a pole of $\chi$ then $U=(1,0)^t$ with eigenvalue $z$. The other eigenvector is $\sigma_z U=(1,0)^t$. Therefore there is a degeneracy and $z^2=1$ and therefore $tr(\calM)=\pm 2$. We can then conclude the following:

\textit{
	The only poles and zeros of the function $\chi$ are generically situated at band edges only, that is, at points where $tr(\calM)=\pm 2$ and $\calM\neq \pm I_2$. They are real when the potential is real.
}

The position of poles and zeros is the clue to understanding the topological properties of the structures. Indeed, a topological transition is characterized by the closing and re-opening of a gap through a continuous variation of the parameters defining the structure. The value of $k_0$ for which the lower and upper bands touch is a critical point of $f(k_0)=\Tr(\calM(k_0))^2-4$, indeed, there it holds $f(z)=0$ and $f'(z)=0$. Therefore: $\calM=\pm I_2$. Since at this point the eigenvectors are $(1,0)^t$ and $(0,1)^t$, this means that a pole and a zero of $\chi$ merge. As a consequence, there is always a pole and a zero at the boundaries of a band gap. In other words, when the topological transition takes place, a pole and a zero change places and therefore they have to merge. We conclude that:

\textit{
	When the potential is real, the topological transitions take place at $\calM=\pm I_2$.
}

Furthermore, the poles and zeros are continuous functions of the parameters and disappear only when they merge. As a consequence:

\textit{
	The sequence of poles and zeros is a topological invariant. It determines the positions of the forbidden and conduction bands
}

This sequence is called the Poles-Zeros pattern.
When losses are added, the situation is more complicated, because in that case the Poles-Zeros pattern is contained in the lower part of the complex plane.
 When the potential is complex, the function $\chi$ is no longer purely imaginary in the conduction bands nor real in the band gaps. It turns out that the function $\chi$ is defined over $F$ and takes values in the projective space, assimilated to the Riemann sphere. This result is immediate whenever the eigenspaces are not degenerated, since the entries of the monodromy matrix are holomorphic.
%
\section{The Berry-Zak phase and the triviality of the bundle}
Let us relate these results to the Berry-Zak phase. This phase is the one that is acquired by a Bloch wave as the Bloch number evolves around the Brillouin zone. It is defined explicitely as follows. Consider the periodic part of a Bloch mode $\psi(k;x)$. The Bloch mode infinitely close to $\psi(k;x)$ is obtained by making a small variation $d k$ and by imposing that $\psi$ is transported without variation at order $1$. To do so, we write that $\psi(k+dk;x)=\psi(k;x)+\partial_k \psi(x;k)dk+O(dk^2)$, and we impose that the variation $\delta \psi$ of $\psi$ belongs to the eigenspace of $\psi$: 
\bq
\delta \psi=\Pi_{\psi}\left(\psi(k+dk;x)-\psi(k;x)\right)
\eq
where $\Pi_{\psi}=| \psi \rangle \langle \psi |$. This gives:
$\delta \psi=\langle \psi,\partial_k \psi \rangle \psi$. This defines the so-called "connection form": $A(k)=\langle \psi,\partial_k \psi \rangle dk$. From this definition, it is seen that $A(k)$ is purely imaginary for a normalized Bloch mode. Indeed, if $\langle \psi,\psi \rangle=1$, then $\langle \partial_k \psi,\psi \rangle+\langle \psi,\partial_k \psi \rangle=0$.

 For a generic Bloch mode $\phi(k;x)=w(k) \psi(k;x)$, the parallel transport of $\phi$ around the Brillouin zone amounts to let $\phi$ evolve in such a way that its variation is orthogonal to the the eigenspace generated by $\psi$ (this is what is done in first order time-independent perturbation theory): ${\langle \delta \phi,\psi \rangle=\partial_k w+A(k) w=0}$ \cite{dewitt}. This gives the differential equation: ${\partial_k w=-A(k)w}$ and therefore:
$w(k)=e^{-\int_{-\pi}^k A(k')} w(-\pi)$. Going around the Brillouin zone defines the so-called Berry-Zak phase $\int_{-\pi}^{\pi} A(k')$ through the integration of the differential equation: $w(\pi)=e^{-\int_{-\pi}^{\pi} A(k')} w(-\pi)$.
A major result obtained in \cite{kohn} and rexpressed in the terms of geometrical phases in \cite{zak} is that, provided the potential as the invertion symmetry, it holds:
$$
e^{-\int_{-\pi}^{\pi} A(k')}=\pm 1.
$$
The Bloch bundle is trivial when the value is $1$ and non-trivial when it is $-1$. This is equivalent to the existence of an \emph{equivariant section}. Here, equivariant means that a section, i.e. a continuous parametrization of a Bloch mode over the Brillouin zone, satisfies a compatibility condition with the inversion symmetry. 

This concept translates easily in our formulation. Indeed, The Bloch modes are represented by the eigenvectors of $\calM(k_0)$ and a conduction band corresponds to an interval of wavenumbers $[k_1,k_2]$. At the boundaries $k_1,k_2$ the monodromy matrix is of one of the following form:
\bq
\begin{pmatrix} \pm 1 & * \\0 & \pm 1 \end{pmatrix} \hbox{ or }
\begin{pmatrix} \pm 1 & 0 \\ * & \pm 1 \end{pmatrix}
\eq
where $*$ is a non-zero element. The corresponding eigenvectors are:
\bq
\begin{pmatrix} 1 \\ 0 \end{pmatrix},\,
\begin{pmatrix} 0 \\ 1 \end{pmatrix}
\eq
The point however is the possibility to follow by continuity an eigenvector around the Brillouin zone. As already said, since the bundle is complex and over $S^1$, such a section necessarily exists \cite{nash}. Here, we request further that it be equivariant, that is, that it be compatible with the group action induced by the inversion symmetry. Specifically, an equivariant section $V(k_0,z)$ should satisfy \bq U(k_0,1/z)=\tilde{\sigma}U(k_0,z), \eq where \bq \tilde{\sigma}=\sigma_z \hbox{ or }\tilde{\sigma}=-\sigma_z. \eq

\begin{itemize}
 \item Assume that we start with the eigenvector $(1,0)^t$ (that is, a pole of $\chi$) for $z=1$. It satisfies the relation: $(1,0)^t=\sigma_z (1,0)^t$. Let us say that we follow the upper part of $S^1$, then we arrive at an eigenvector $(a,b)^t$ for $z=-1$ that corresponds either to a zero or to a pole. Going around the lower part, we arrive at the eigenvector $(a',b')^t$ for $z=-1$. Keeping into account the equivariance of the section, we impose that $(a',b')^t=\sigma_z (a,b)^t$, that is: $a=a',\, b=-b'$. For the section to be continuous, we have of course to impose: $a=a',\, b=b'$, but then, of course, $b=0$. Therefore we end up with the eigenvector $(1,0)^t$. We have therefore fulfilled the conditions for the existence of a section.
\item Let us start now with a zero $(0,1)^t$. This does fullfill the requirement of equivariance provided that we write $(0,1)^t=-\sigma_z(0,1)^t$. If there is also a zero at $z=-1$, the same gluing works to provide a global section.

\item The situation is different if we start with a zero $(0,1)^t$ and end with a pole $(1,0)^t$. This time we start with the condition $(0,1^t)=-\sigma_z (0,1)$ and end with the condition $(1,0)^t=\sigma_z(1,0)^t$. Therefore the section is not globally equivariant and it has to be twisted.
\end{itemize}
We end up with the following conclusions:
\begin{enumerate}
\item The pole-pole or zero-zero cases correspond to a Berry phase equals to $0$,
\item the pole-zero or zero-pole cases correspond to a Berry phase equals to $\pi$. 
\end{enumerate}

We note that following our approach, the Berry-Zak phase is very easy to compute as it suffices to consider the form of the monodromy matrix at the boundaries of the conduction bands.
\section{The bulk-boundary correspondence}
\begin{figure}
	\begin{center}
		\includegraphics[width=12cm]{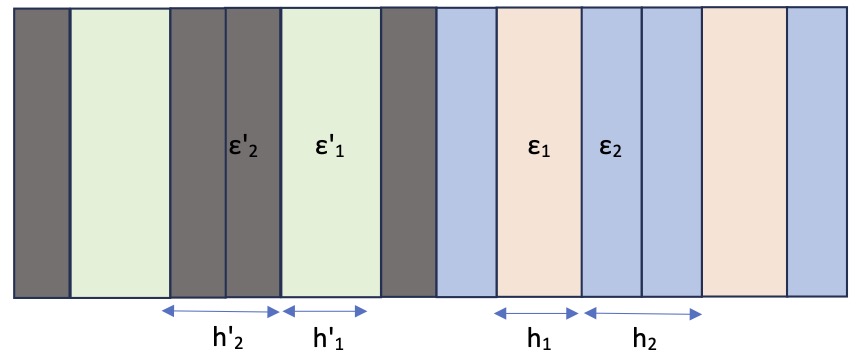}
		\caption{The structure made with two 1D photonic crystals with different topological properties.}
		\label{2media}
	\end{center}
\end{figure}

Let us consider a structure made of two semi-infinite photonic crystals put side by side such as depicted in fig.(\ref{2media}), characterized by the monodromy matrices $\calM_1$ and $\calM_2$. Our point is to investigate under what conditions it can exist a boundary mode and how it can be characterized topologically by means of the properties of the $\chi$ function.
%
%
%
%

The edge states are ruled by the following result.
Assume the photonic crystal defined by $\calM_1$ extends over $x>0$ and that defined by $\calM_2$ over $x<0$.
At $k_0$ the eigenvalues of $\calM_1$ and $\calM_2$ are real and of the form $(z_1,1/z_1)$ and $(z_2,1/z_2)$. Let us assume that $|z_{1,2}|<1$. A boundary mode is characterized by its initial value $U$ at the junction between the media. For the mode to be bounded, the vector $U$ should be damped along $x>0$ and $x<0$, therefore it should hold
\bq 
\calM_1 U=z_1 U \hbox{ and }\calM_2 U=1/z_1 U.
\eq
 This means first that, generically, $\calM_1$ and $\calM_2$ have a common set of eigenvectors, hence these matrices commute. Second, because of the symmetry of the potential, the second eigenvector is $V=\sigma_z U$

\textit{
Let $\calM_1$ and $\calM_2$ be the monodromy matrix of each photonic crystal. For a Bloch wavevector $k_0$, there exists an eigenvector vector $U(k_0)$ defining an edge state provided the following conditions are fulfilled
	\begin{itemize}
		\item The matrices $\calM_1(k_0)$ and $\calM_2(k_0)$ have a common gap at the Bloch wavevector $k_0$, i.e. $|\mathrm{tr}(\calM_1)| >2$ and $|\mathrm{tr}(\calM_1)| >2$,
		\item the matrices $\calM_1(k_0)$ and $\calM_2(k_0)$ commute:$[\calM_1(k_0),\calM_2(k_0)]=0$,
		\item the associated Bloch functions $\chi_1(k_0)$ and $\chi_2(k_0)$ have opposite signs.
	\end{itemize}
}
The last two conditions are equivalent to the single following one:

\textit{
The Bloch functions $\chi_1(k_0)$ and $\chi_2(k_0)$ are opposite: $\chi_1(k_0)=-\chi_2(k_0)$.
}

The link with the Poles-Zeros pattern can now be deduced. Recall that at the boundaries of the gaps the functions $\chi$ necessarily have a zero or a pole and that they have a constant sign within a band gap. This means that, provided $\chi_1$ and $\chi_2$ have opposite signs in a band gap, when the pattern inside a gap is Pole-Zero for one structure and Zero-Pole for the other, the functions $\chi_1$ and $-\chi_2$ necessarily cross and there necessarily is an edge mode.


A few words are in order as to the fact that the existence of a mode is linked in a very strict way to the symmetry property of the potential, as far as the theoretical analysis is concerned. However, an edge mode inside a gap corresponds to a pole of the scattering matrix (i.e. of the reflection and the transmission coefficients). Since we expect the pole to be a continuous function of the parameters it seems paradoxical that it could disappear suddenly when the symmetry is broken since it can be broken in a continuous fashion by moving continuously the origin of the basic cell. To resolve this apparent paradox, one should recall that the reflection and transmission coefficients are defined for two finite structures put side by side (see \cite{dofresnel} for the definition of the reflection coefficient for a semi-infinite structure). When two semi-infinite structures are considered, the mode should be evanescent in both structures away from the edge and this is a strict condition. When finite structures are considered (containing each N periods), and with an plane wave incident field, there necessarily are anti-evanescent waves in addition to evanescent waves in order to fullfill the boundary conditions. Therefore, breaking the symmetry condition does not kill suddenty the edge mode, rather its proper wavenumber is shifted towards the edges of the band gap and, as the number of periods N tends to infinity, the edge mode disappears continuously by being absorbed at the edges of the band gap.

\begin{figure}
	\begin{center}
		\includegraphics[width=12cm]{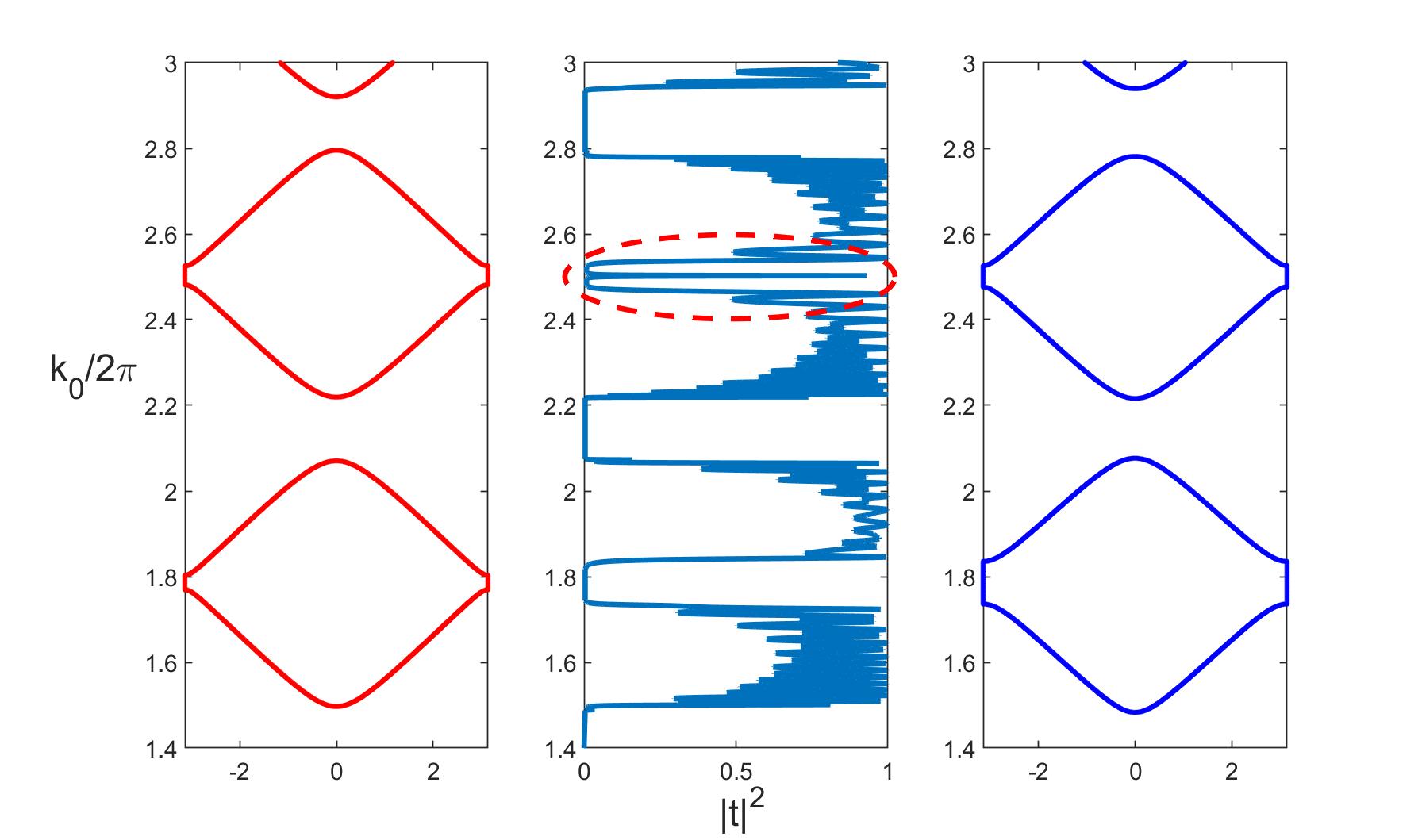}
		\caption{band structures for two photonic crystals with parameters $\ep_1=3.8,\,\ep_2=1$, $h_1=0.42,\,h_2=0.58$,on the left, and $\ep_1=4.2,\,\ep_2=1$, $h_1=0.38,\,h_2=0.62$, on the right. The origin is chosen is such a way that the basic cell contains $3$ layers of width $h_1/2,h_2,h_1/2$ and permittivities $\ep_1,\ep_2,\ep_1$. In both cases the permeabilities are equal to $1$. In the middle the transmission spectrum is given. It is obtained with a finite structure comprising 10 periods of each photonic crystal.}
		\label{sidebyside}
	\end{center}
\end{figure}

Let us illustrate these results numerically. The two band structures corresponding to each photonic crystal are given in fig. (\ref{sidebyside}).
A finite structure made of 10 periods of each photonic crystal is considered. The transmission spectrum for an incident plane wave is plotted in fig. (\ref{sidebyside}) The existence of an edge state is detected as a peak inside the band gap for the forbidden band situated inside the interval of $k_0/2\pi \in [2.4, 2.6]$. 
\begin{figure}
	\begin{center}
		\includegraphics[width=12cm]{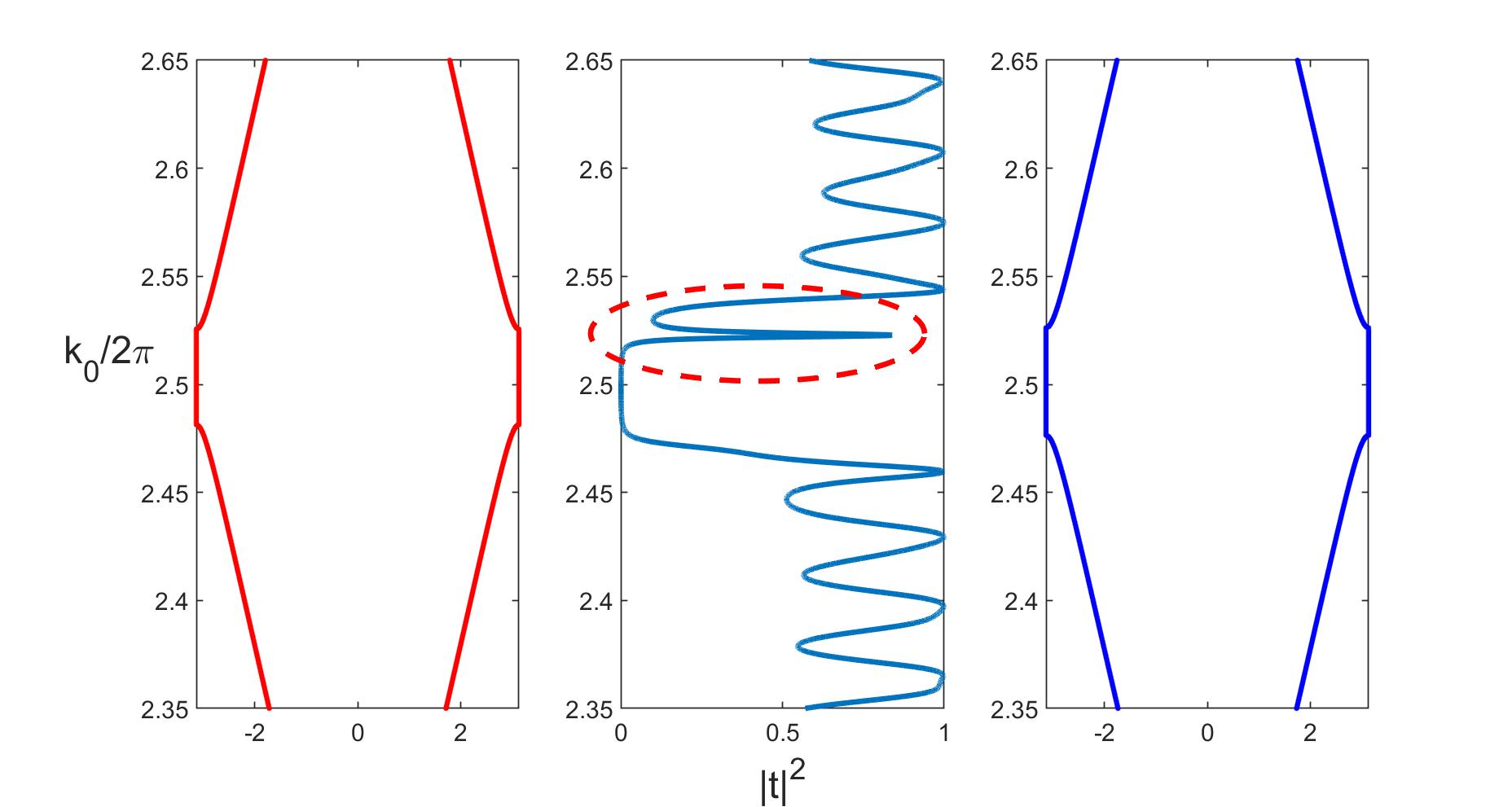}
		\caption{band structures for the same photonic crystals as in fig. (\ref{sidebyside}) but the photonic crystal on the left no longer satisfies the symmetry condition. The formula for the widths of the layers of the basic cell is now $3/5h_1,h_2,2/5h_1$ and the permittivities are $\ep_1,\ep_2,\ep_1$.On the right, the transmission spectrum is given. An edge mode is still present but it has moved towards the upper edge of the bandgap.}
		\label{zoombroken}
	\end{center}
\end{figure}

\begin{figure}
	\begin{center}
		\includegraphics[width=12cm]{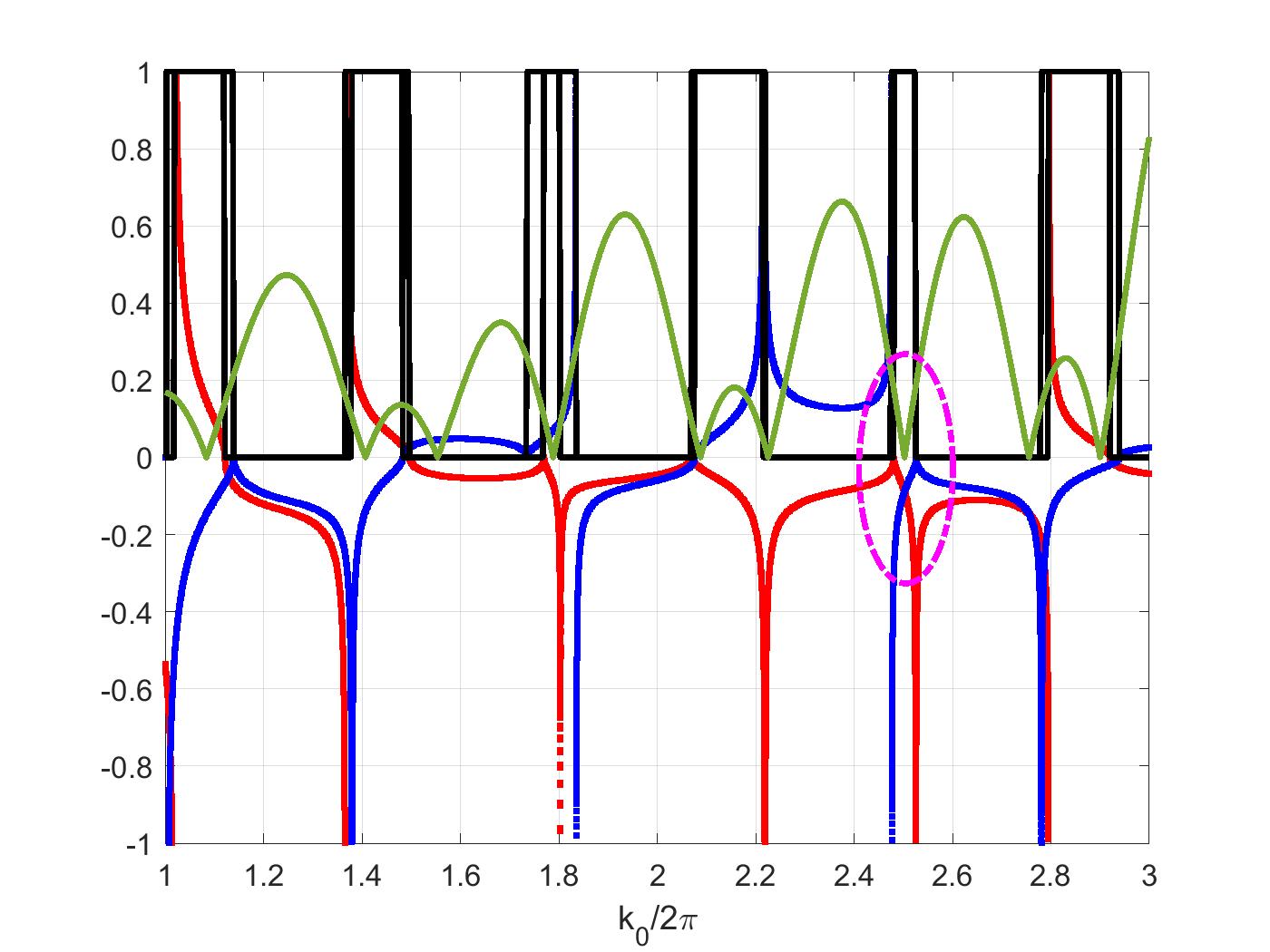}
		\caption{In blue the function $\chi_1$, in red the function $-\chi_1$. The commutator of $\calM_1$ and $\calM_2$ is plotted in green. The positions of the band gaps are indicated by the black indicator (value $1$ in the band gaps). The edge mode appears when $\chi_1$ and $-\chi_2$ cross and the commutator is null.\label{curvechi}}
	\end{center}
\end{figure}

\begin{figure}
	\begin{center}
		\includegraphics[width=12cm]{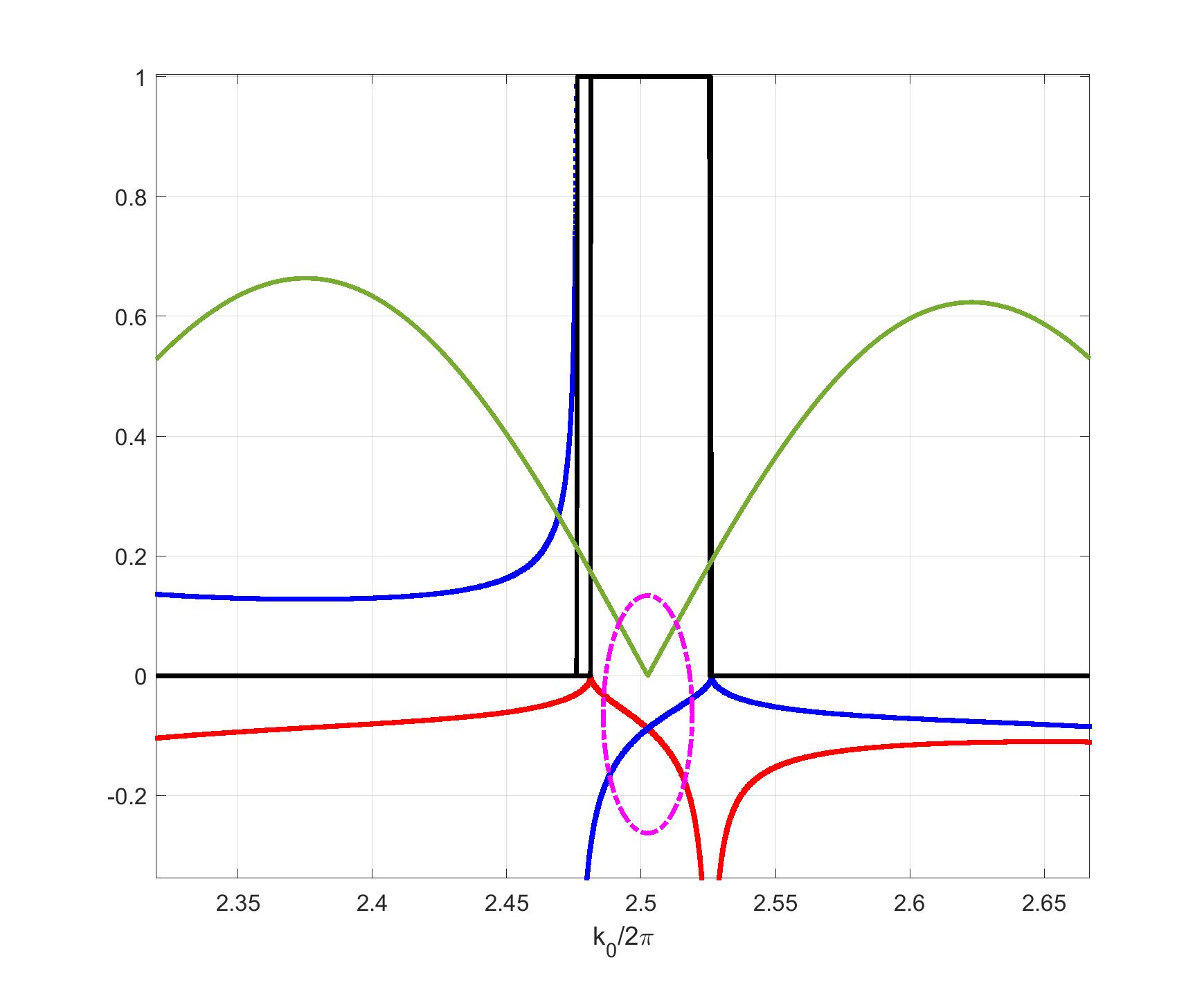}
		\caption{Same as fig. (\ref{curvechi}) but zoomed in.\label{curvechizoom}}
	\end{center}
\end{figure}

\begin{figure}
	\begin{center}
		\includegraphics[width=8cm]{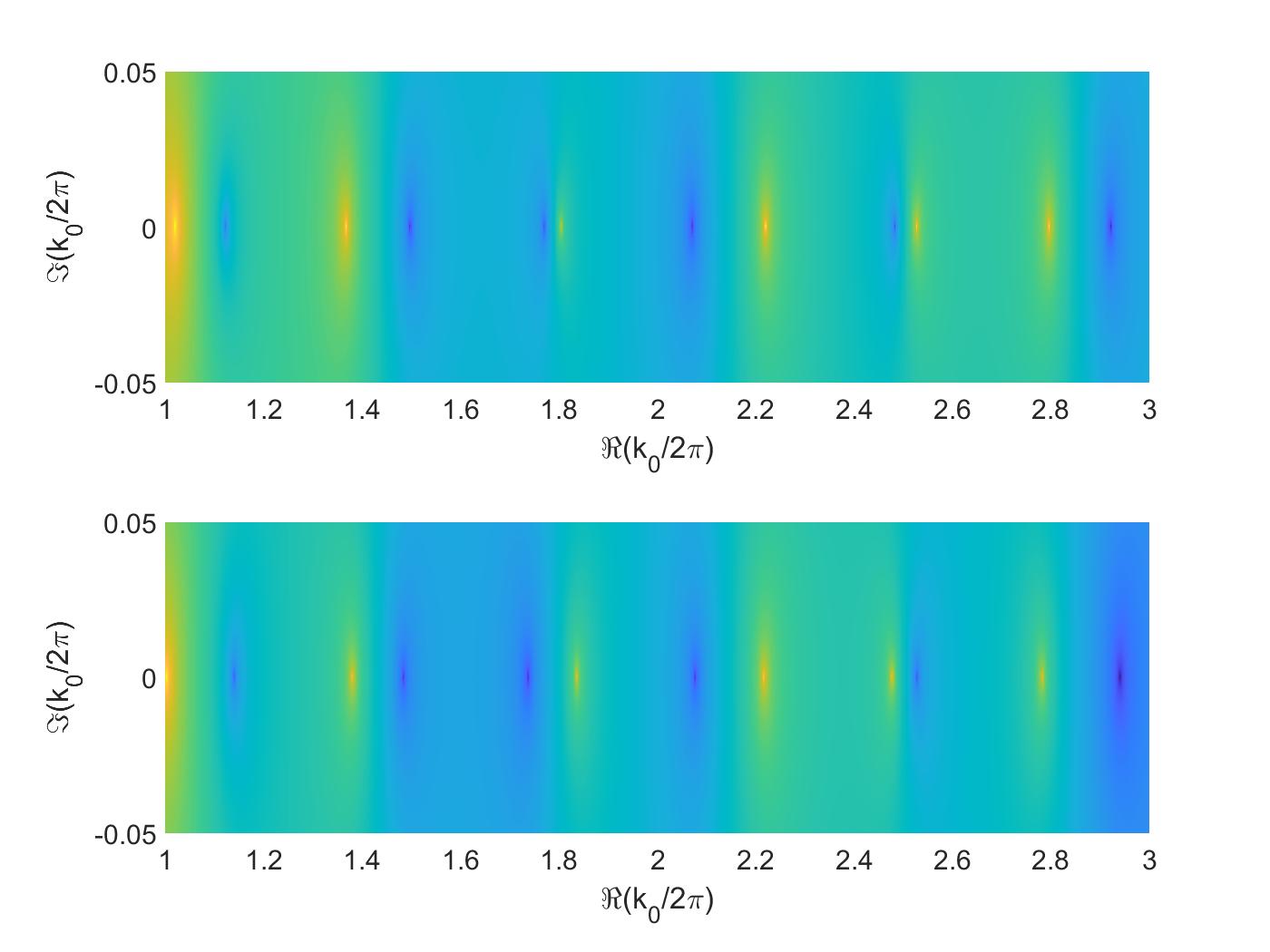}
		\caption{Poles-zeros patterns for the two structures described in fig. (\ref{sidebyside}). The bright spots are the poles and the blue ones the zeros. Around the value $k_0/2\pi=2.5$ one of the structure has a Zero-Pole pattern while the other one has a Pole-Zero pattern.\label{polezero}}
	\end{center}
\end{figure}

In fig. (\ref{curvechi},\ref{curvechizoom}), we have plotted $\chi_1$ (in red) and $-\chi_2$ (in blue), as well as the commutator of $\calM_1$ and $\calM_2$ in green. The position of the band gaps is indicated in black. As can be seen on fig. (\ref{curvechi},\ref{curvechizoom}), the edge state correspond indeed to a value of $k_0$ for which the functions $\chi_{1,2}$ cross and the commutator of the monodromy matrices is null.
Let us now see what happens when complex values of $k_0$ are used. In fig. (\ref{polezero}), we plot the absolute value of $\chi$ in the complex strip around the real axis. The poles are indicated by bright spots and the zeros by blue spots.
We see clearly that the structures have the same Poles-Zeros pattern, except for the gap around $k_0/2\pi=2.5$ where one of the structure has the pattern Zero-Pole and the other one the pattern Pole-Zero. In this band gap, there is a boundary mode.
In fig. (\ref{zoombroken}), we have plotted the band structure and transmission spectrum when the right-hand side photonic crystal does not fullfill the symmetry condition that $V(-x)=V(x)$. As explained in the discussion above, the edge mode is still present but is shifted towards the boundary of the band gap.

\begin{figure}
	\begin{center}
		\includegraphics[width=12cm]{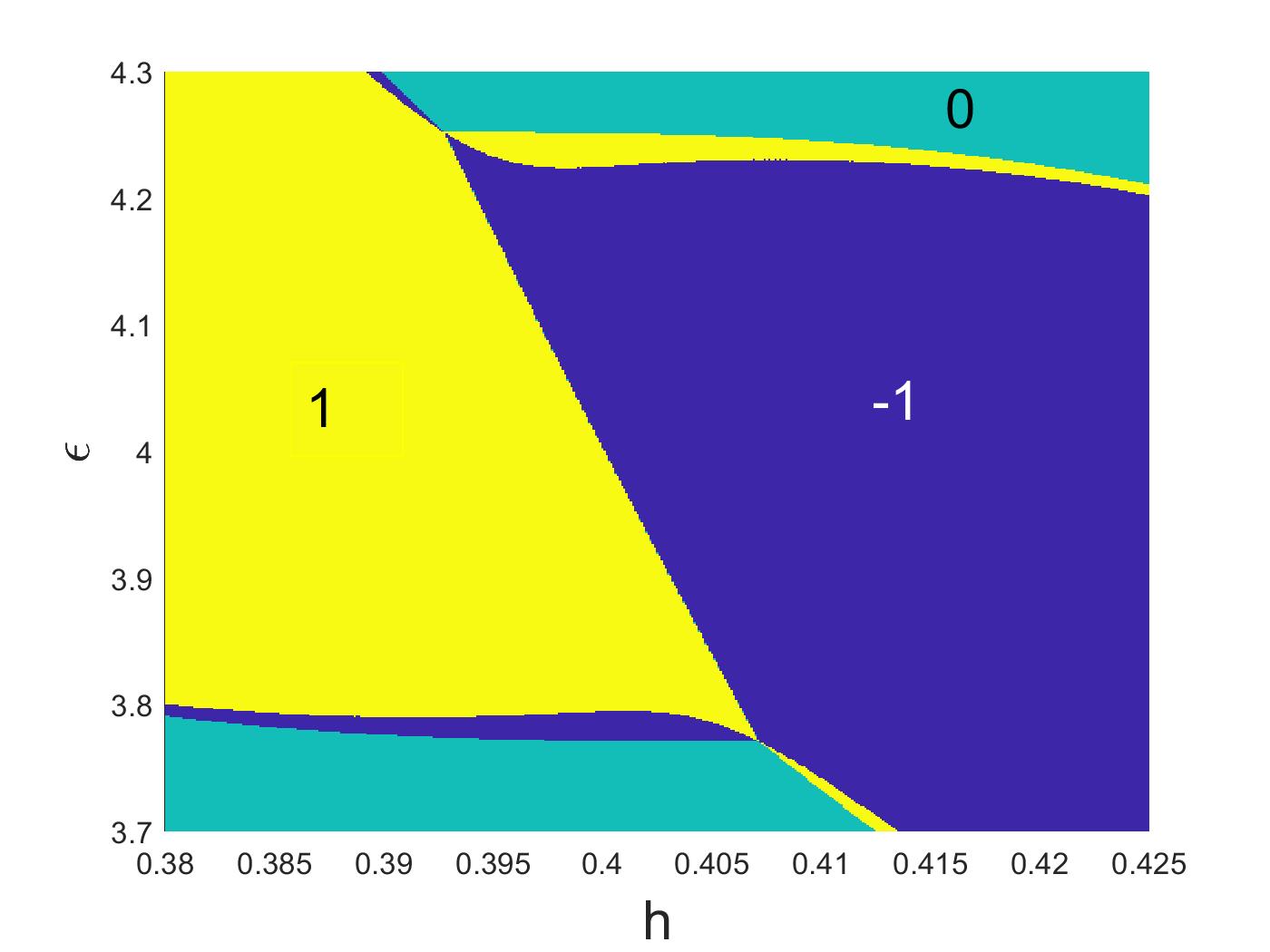}
		\caption{Phase diagram for a stratified medium with two slabs. The parameters are the permittivities $\ep_{1,2}$ and the widths $h_{1,2}$ of the slabs. Since the period is $1$, there is only one width parameter since $h_1+h_2=1$. The diagram corresponds to only a band gap in the interval $[2.4; 2.6]$. The index $1$ is attributed to the pattern Pole-Zero and the index $-1$ to the pattern Zero-Pole. The index $0$ is for structures which do not have a band gap in the considered interval. \label{phasediagram}}
	\end{center}
\end{figure}

By varying the values of the permittivities and the width of the layers, it is possible to obtain a phase diagram for the topological properties, i.e. the pole and zero (cf. fig. (\ref{phasediagram})). It can be seen that the regions of interest are separated by lines corresponding to the closing of the band gap.

\begin{figure}
	\begin{center}
		\includegraphics[width=12cm]{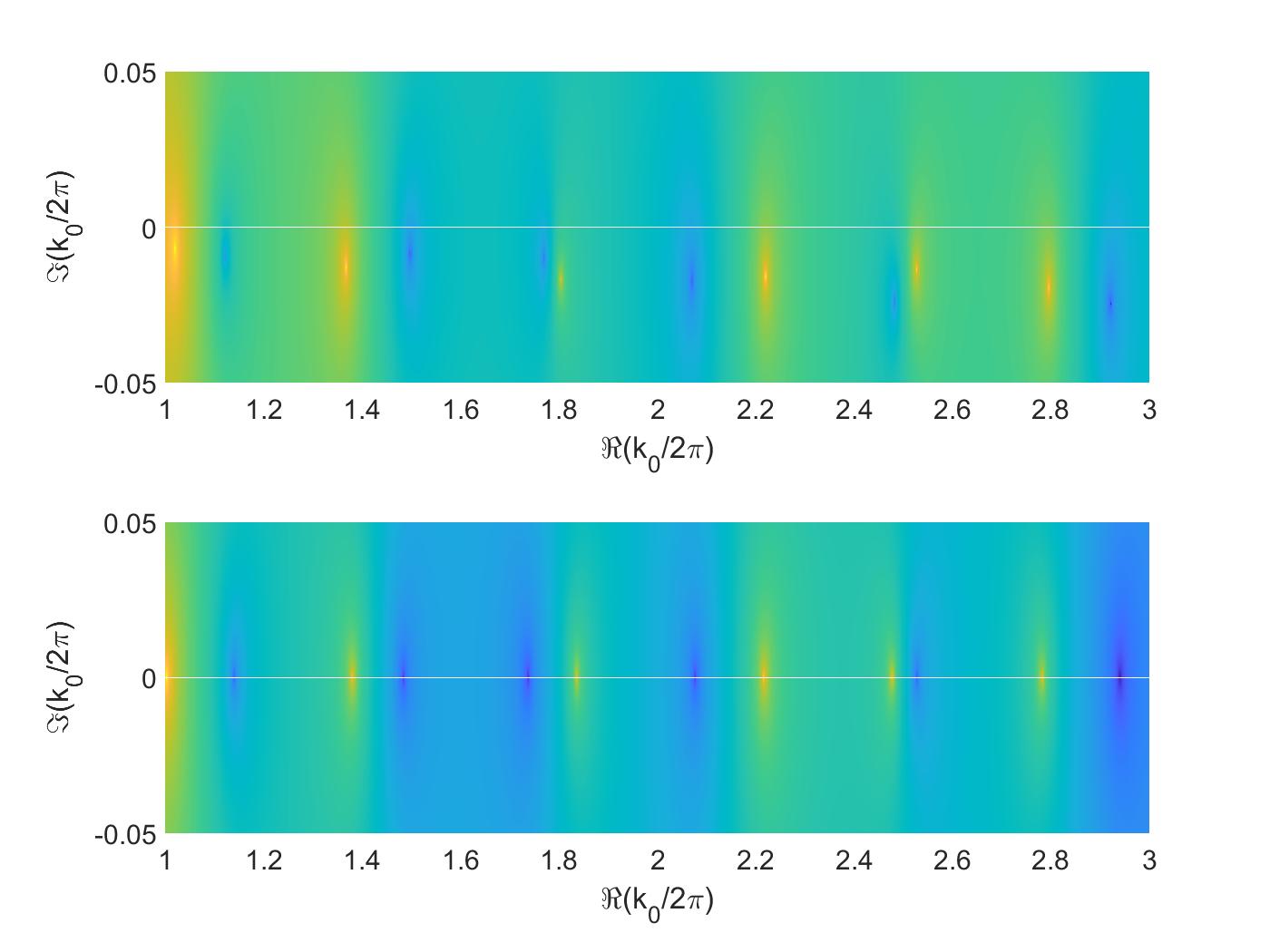}
		\caption{Same as fig. (\ref{polezero}) but losses have been added to the structure corresponding to the upper Poles-Zeros pattern. The other structure remains lossless. The Poles-Zeros pattern is still there, except that the poles and zeros have moved towards the lower part of the complex planbe of wavevectors. It can still be seen that the band gap around $k_0/2\pi=2.5$ corresponds to a Pole-Zero/Zero-Pole pattern.\label{poleperte}}
	\end{center}
\end{figure}

Our point is now to study the situation where the potential is complex, that is, an imaginary part is added to the permittivities. The new Poles-Zeros pattern is given in fig.(\ref{poleperte}).

\begin{figure}
	\begin{center}
		\includegraphics[width=8cm]{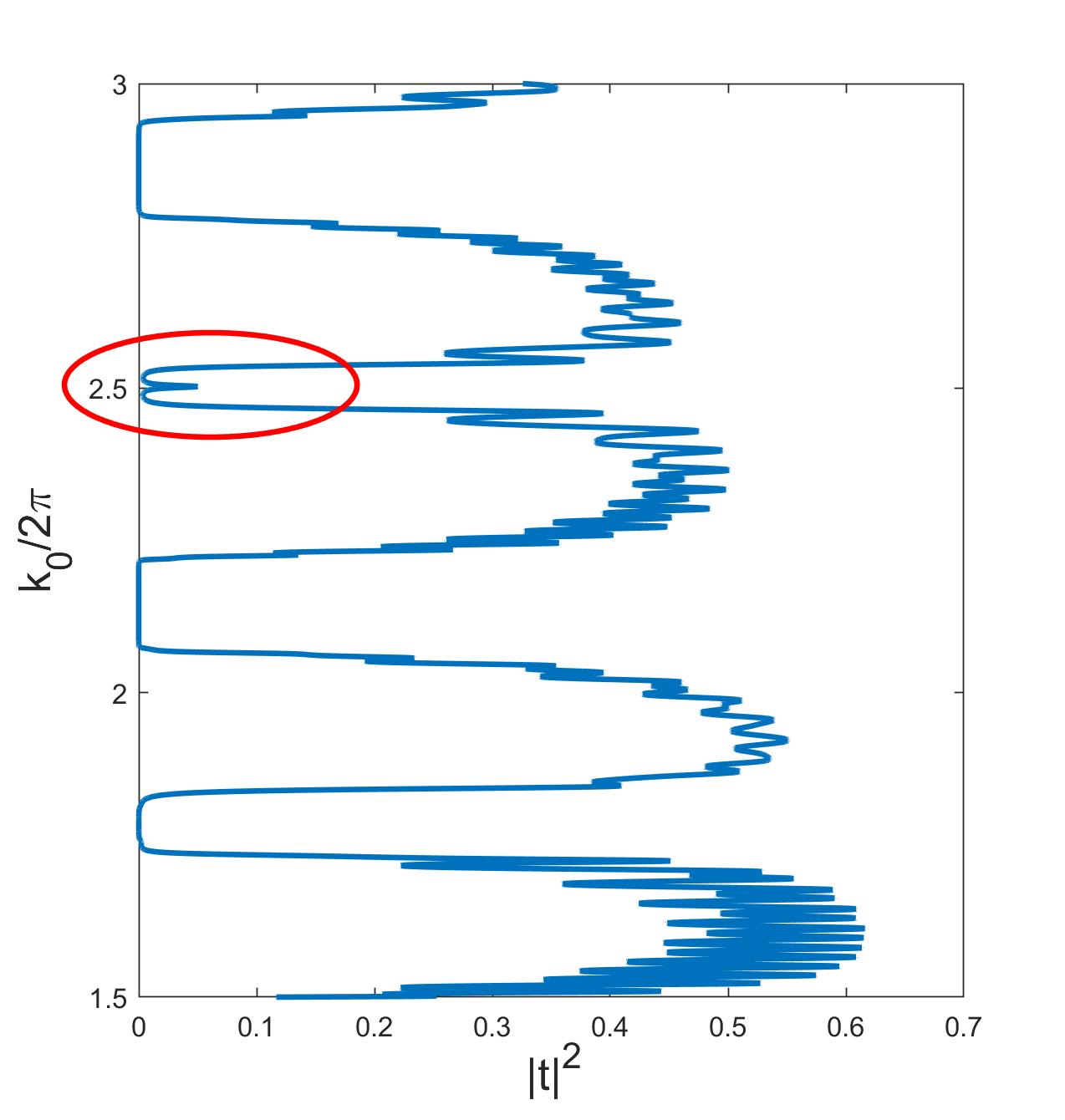}
		\caption{Transmission spectrum for the finite structure made of two structures side by side. Losses are present but there is still an edge mode, that appears as a peak inside a band gap. Due to losses, the peak is larger and the maximum value is smaller.\label{transloss}}
	\end{center}
\end{figure}

There, it can be seen that the poles and zeros have moved towards the lower part of the complex plane of wavevectors. Still the pole-zero structure is preserved. There is a continuous transformation, mathematically speaking a homotopy, between the poles and zeros structures of both photonic crystals. When a pole and zero exchange places by continuously varying a parameter, there is a configuration for which the pole and the zero have the same real part. This amounts to saying that the gap closes there. Consequently, as far as the Poles-Zeros pattern remains close to the real axis, these patterns continue to characterize topologically the structures. Indeed, when plotting the transmission spectrum for the finite structure with losses (cf. fig. (\ref{transloss})), a mode can be seen to remain (as an enlarged peak) in the same band gap. Of course, the life-time of the mode is now much shorter due to material losses added to radiative losses.
\section{Conclusion}
It is customary to analyze the topological properties of one dimensional structures by using the familiar concept of Zak phase, directly linked to Berry's connection. We have shown here that another approach can be put forward, by using the poles and zeros of a function defined for all energies and not only for that corresponding to propagating modes. By using this tool, the extension to the classification of media with losses, or non-hermitian problems, is straightforward and avoids the difficulties encountered when trying to extend the Berry connection approach to non-hermitian system. More generally, the proposed approach gives possible theoretical insights for handling the case of complex energies and for analysing the Bloch variety \cite{gieseker}.

\end{document}